# "Smart Girls" versus "Sleeping Beauties" in the Sciences: The Identification of Instant and Delayed Recognition by Using the Citation Angle

Fred Y. Ye [a, b*] Lutz Bornmann [c]

[a] *School of Information Management, Nanjing University, Nanjing 210023, CHINA*

[b] *Jiangsu Key Laboratory of Data Engineering and Knowledge Service, Nanjing 210023,CHINA*

[c] *Administrative Headquarters of the Max Planck Society, Division for Science and Innovation Studies, Hofgartenstr. 8, D-80539 Munich, GERMANY*

[*] Corresponding author: yye@nju.edu.cn

**Abstract**: In recent years, a number of studies have introduced methods for identifying papers with delayed recognition (so called "sleeping beauties", SBs) or have presented single publications as cases of SBs. Most recently, Ke et al. (2015) proposed the so called "beauty coefficient" (denoted as B) to quantify how much a given paper can be considered as a paper with delayed recognition. In this study, the new term "smart girl" (SG) is suggested to differentiate instant credit or "flashes in the pan" from SBs. While SG and SB are qualitatively defined, the dynamic citation angle $\beta$ is introduced in this study as a simple way for identifying SGs and SBs quantitatively – complementing the beauty coefficient B. The citation angles for all articles from 1980 (n=166870) in natural sciences are calculated for identifying SGs and SBs and their extent. We reveal that about 3% of the articles are typical SGs and about 0.1% typical SBs. The potential advantages of the citation angle approach are explained.

## 1. Introduction

The long-term success of publications measured in terms of citations is suggested to be an important indicator in research evaluation (Leydesdorff et al., 2016). However, in the current practice of research evaluation, relatively short citation windows (up to three or five years) are normally used (Waltman, 2016; Wang, 2013). One important reason for the use of short citation windows in the evaluative context is that reviewers are generally interested in the performance over recent years. Both time perspectives in research evaluation – long- and short-term citation impact – also produce specific citation phenomena, which can be called delayed recognition and instant credit.

Whereas gaining instant credit – i.e. when most of the credit arises within three to five years after publication – is the common citation distribution of many papers having a considerable number of citations, delayed recognition is a more or less rare phenomenon. In recent years, a number of studies have introduced methods for identifying papers with delayed recognition or have presented single publications as cases of papers with delayed recognition. Most recently, Ke et al. (2015) proposed the so called "beauty coefficient" (denoted as B) to quantify the extent to which a given paper can be considered as a paper with delayed recognition. Since B is an interesting parameter-free coefficient, which can be used to identify sleeping beauties (SBs) among publications published in different fields and publication years, we would like to take up and extend the coefficient with the instant credit perspective.

Following a brief literature review on SBs, we will present the extended coefficient for unifying instant credit and delayed recognition.

## 2. Literature review

Since 1955, when Garfield (1955, 1979) introduced the citation index, citation analysis has become more and more popular. Over the last few decades citation analysis has developed into the core method in scientometrics where two research topics are in the focus: the development and improvement of citation-based indicators as well as the investigation of specific citation phenomena. A few examples of the development and improvements of indicators are the h-index (Hirsch, 2005), citation percentiles (Bornmann &Mutz, 2011), the integrated impact indicator (I3, Leydesdorff & Bornmann, 2011), and matrix metrics (Ye & Leydesdorff, 2014).

At the end of the 1980s, Garfield (1989a, 1989b, 1990) started to investigate the delayed recognition phenomenon of publications in some research fields. He chose two criteria to address delayed recognition: 1) highly cited papers which had low citation counts for the first 5 or more years (with more than 10 years being preferred); 2) Papers with a low initial citation frequency, i.e. being close to the average of one citation per year. By using both criteria, he found 5 typical examples of delayed recognition from the Science Citation Index (SCI) Citation Classics via annual citation curves, including Steven Weinberg's paper published in 1967 (Weinberg, 1967).

The papers of Garfield (1989a, 1989b, 1990) were starting points for further studies on delayed recognition and possible methods for detecting corresponding papers. Glänzel et al. (2003) defined a paper published in 1980 as having delayed recognition, if it received (a) only one citation in an initial 3-year period or (b) at most two citations in an initial 5-year period, and (c) if it is now highly cited (i.e. has received at least 100 citations in the remaining period up to 2000). The authors found 77 papers out of the almost 450,000 publications under the weak condition (a) and 29 papers under the stronger condition (b). After changing condition (c) to papers with at least 50 citations and 10 times the average journal impact, the selection resulted in a set of 60 papers (under a, the weak condition) and 16 papers (under b, the strong condition). The 3- or 5-year citation window was chosen, because – as a rule – more than 80% of the papers in a database are cited in an initial 3-year citation window and more than 90% in an initial 5-year citation window in terms of first-citation statistics (Glänzel et al., 2003).

In a short paper by Glänzel and Garfield (2004), delayed recognized papers are defined as those papers which were initially rarely cited during a period of 5 years but became highly cited during the next 15 years (at least 50 citations or 10 times the journal's 20-year cumulative impact factor). After analyzing the citation histories of 450,000 articles and reviews indexed in the 1980 edition of the SCI, they found that a statistically marginal share of 1.3 per 10,000 papers were neglected initially, but generated relatively high citation impact later on.

van Raan (2004) was the first to name publications with delayed recognition as SBs. Since then, this term has become established in scientometrics. The back-story for this naming is as follows: a SB is a princess (a single publication) which sleeps (goes unnoticed) for a long time, is awakened rather suddenly by a prince (another publication) and generates a lot of citation impact after that. Quantitatively, van Raan (2004) suggested three variables for identifying SBs: (1) Depth of sleep ($C_s$): the publication received at most 1 citation per year on average (deep sleep) or between 1 and 2 citations per year on average (less deep sleep) for a few years after publication. (2) Length of sleep ($s$): the duration of the sleeping period. (3) The awakening intensity ($C_w$): the number of citations per year during the 4 years following the sleeping period. As an important contribution, van Raan (2004) developed the so called Grand SB Equation: $N=f\{s, c_s, c_w\} \sim s^{-2.7} \cdot c_s^{2.5} \cdot c_w^{-6.6}$, where $N$ is the number of SBs.

Costas et al. (2010) technically defined various types of paper durability in terms of citations regardless of publication year and total number of citations. In their study, "Yr 50%" identifies the year when a publication has received at least 50% of its citations; "P25" denotes the prior 25% citations as a one-fourth quartile criterion; and "P75" denotes the prior 75% citations as a three-fourth quartile criterion. Costas et al. (2010) concluded that "Yr 50%" < "P25" characterizes instant credit (in other words "flashes in the pan" or "early rise, rapid decline") (Aversa, 1985). "Yr 50%" > "P75" identifies delayed recognition (in other words "delayed rise, no decline"); and "P25" < "Yr 50%" < "P75" indicates "normal" or "medium rise-slow decline".

Li and Ye (2012) introduced the "all-elements-sleeping-beauties" (ASBs), where "spindles, sleeping beauties, and princes" co-exist. The citation distribution of ASB papers follows the story in the fairy tale: initially, the papers generate considerable impact as if the SBs are not asleep at the beginning. Then, they start a low citation period as if the SBs prick their fingers on the spindles and fall into a sleeping period. This period is finished by publication years with considerable impact again. In a follow-up paper, Li et al. (2014) defined the "heartbeat spectra" of SBs. Whereas the "heartbeat" of a SB is its annual citations in the sleeping period, the "heartbeat spectrum" is the vector of the SB's heartbeat. Let $c_i$ denotes the number of citations which the SB received in the $i^{th}$ year of the sleeping period. Then, the SB's heartbeat in the $i^{th}$ year is $c_i$. Vector H = ($c_1, \ldots, c_i \ldots, c_n$) becomes the heartbeat spectrum, where n indicates the duration of the sleeping period. Two further studies (Huang et al., 2015; Li & Shi, 2016) deal with the awakening of SBs.

The most recent publication which introduced a method for the identification of SBs is Ke et al. (2015). The beauty coefficient B is defined as follows (for the purpose of simplifying, we use $c_m$ instead of $c_{tm}$):

$$B = \sum_{t=t_0}^{t_m} \frac{\frac{c_\mathrm{m} - c_0}{t_m} t + c_0 - c_t}{\max\{1, c_t\}} \qquad (1)$$

where $c_t$ is the number of citations received in the $t^{th}$ year after publication and $t$ the age of a paper. A paper receives the maximum number $c_m$ of annual citations at time $t_m$. The equation of the straight line (l) is

$$l : c = \frac{c_\mathrm{m} - c_0}{t_m} t + c_0 \qquad (2)$$

This line connects two points $(0, c_0)$ and $(t_m, c_m)$ in the annual citation curve.

According to Cressey (2015), the beauty coefficient B is an elegant and effective method for the identification of SBs in big data. In this paper, we try to reveal its dynamic characteristics and extend B – the SB coefficient – by a "smart girl" (SG) component. Furthermore, a simple indicator – the citation angle – is suggested for unifying the approaches of instant and delayed recognition.

With the distinction between SBs and SGs in this study, we follow Baumgartner and Leydesdorff (2014) who define two groups of papers: (1) Papers which have a lasting impact on a field. These papers can be called "citation classics" or "sticky knowledge claims". SBs are a specific sub-group among these papers, because the lasting impact is not combined with (substantial) citation impact shortly after publication. (2) The other paper group defined by Baumgartner and Leydesdorff (2014) has an initial burst of citation impact followed by a fast decrease shortly after publication. According to Baumgartner and Leydesdorff (2014) these papers are contributions at the research front and can be defined as "transient knowledge claims". Comins & Leydesdorff (2016) used the method called "multi-References Publication Years Spectroscopy" (multi-RPYS – a method which is based on the analysis of cited references in a paper set) to investigate the existence of both paper types empirically.

## 3. Methodology

### 3.1 Definitions

*Definition of "smart girl" (SG), "sleeping beauty" (SB), and zero point*

A typical SG (or "flash in the pan") is defined as a publication with an early citation peak and later annual citations which are much lower than the early peak. A typical SB ("delayed recognition") is defined as a publication with a late citation peak, and prior annual citations which are much lower than the peak citations. The year of the zero point is set to 1 year subsequent to the publication year and (0, 0) is set to the zero point. At the zero point, both time and citation number are 0. According to these three definitions (SG, SB, and zero point), a unified analytical framework for SGs and SBs can be introduced. In Figure 1, $\beta$ is a dynamic angle between the straight line (linking the zero point with the citation peak) and the time axis. If the zero point were not set to (0,0), the line $l$ in the figure would not be a straight line and the citation angle could not be formed.

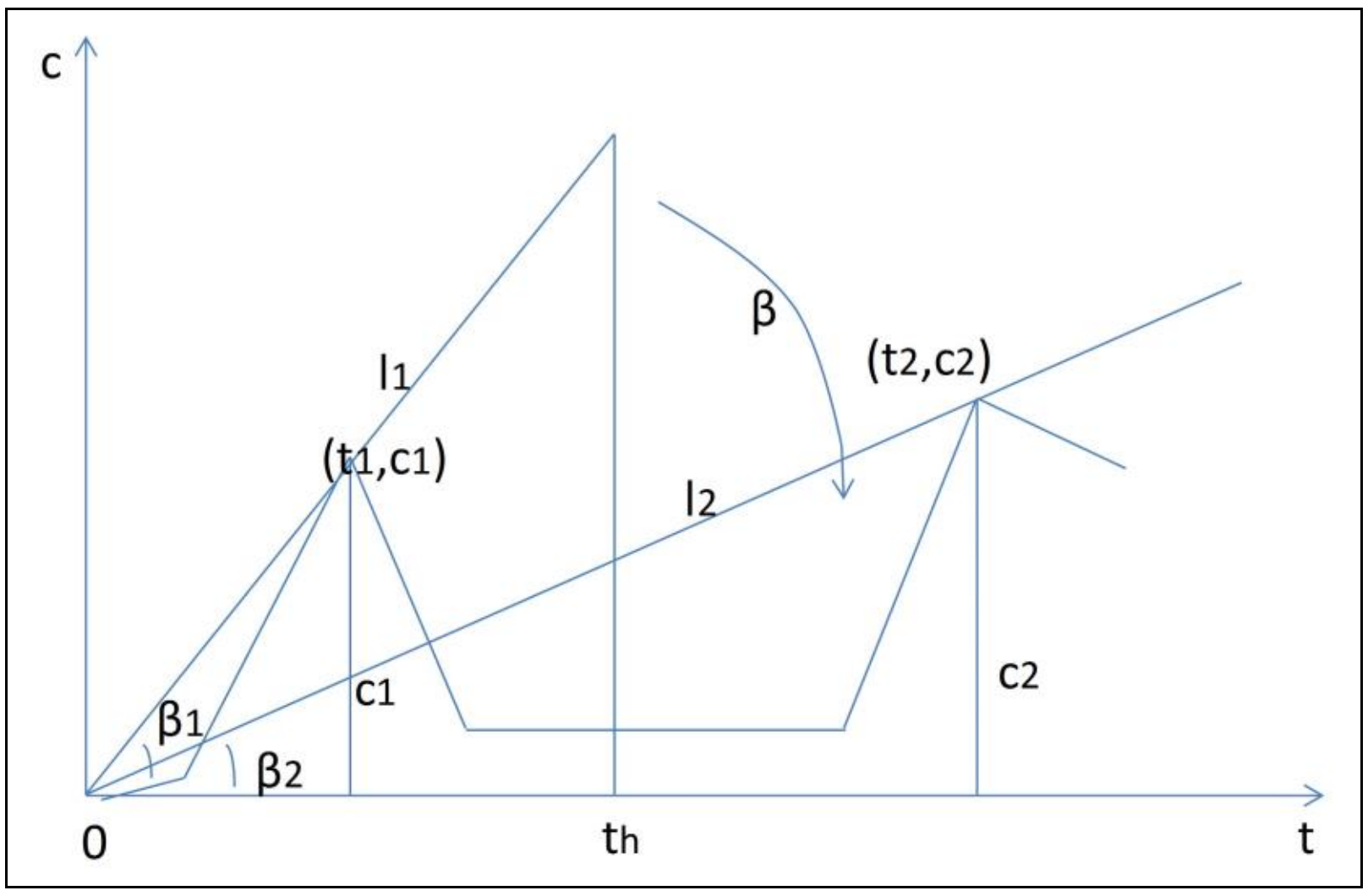

Figure 1. The unified analytical framework for "smart girls" (SGs) and "sleeping beauties" (SBs)

*Definition of the citation angle*

The straight line $l$ links the zero point with any citation peak in the annual citation distribution (see Figure 1). Citation angle $\beta$ is the angle between the straight line and the time axis. As citation distributions can be characterized by many citation peaks, $\beta$ is a dynamic angle with many possible linking lines. If the line is fixed, the angle is unique.

Suppose the straight line $l_1$ connects two points $(0, 0)$ and $(t_1, c_1)$ in the annual citation distribution. Then, the equation of $l_1$ is

$$l_1 : c = \frac{c_1}{t_1} t = \tan\beta_1 \cdot t \qquad (3)$$

where $\beta_1$ is suggested to be the first or early citation angle or "smart" angle.

Similarly, suppose the straight line $l_2$ connects two points ($0$, $0$) and ($t_2$, $c_2$) in the annual citation distribution, the equation of $l_2$ is

$$l_2 : c = \frac{c_2}{t_2} t = \tan\beta_2 \cdot t \qquad (4)$$

where $\beta_2$ is suggested to be the second or late citation angle or "sleeping" angle.

As the zero point is set up with keeping $c_0$ as the citation symbol in the publication year, the equation of the straight line of Eq. (2), connecting ($0$, $0$) and the maximum peak ($t_m$, $c_m$) with the citation angle $\beta_m$, becomes

$$l : c = \frac{c_\mathrm{m}}{t_m} t = \tan\beta_\mathrm{m} \cdot t \qquad (5)$$

Thus, the beauty coefficient can be simplified as B' linking with citation angle as

$$B' = \sum_{t=0}^{t_m} \frac{\frac{c_\mathrm{m}}{t_m} t - c_t}{\max\{1, c_t\}} = \sum_{t=0}^{t_m} \frac{\tan\beta_m \cdot t - c_t}{\max\{1, c_t\}} \qquad (6)$$

There are $\max\{1,c_h\}=c_1$, $\max\{c_h,c_t\}=c_2$, and $c_m=\max\{1,c_t\}=\max\{c_1,c_2\}$ (i.e. $\beta_m= \max\{\beta_1, \beta_2\}$), while $t$ is the total citation time.

### 3.2 Criteria for identifying "smart girls" (SGs), "sleeping beauties" (SBs), and "all-elements-sleeping-beauty" (ASBs) in citation data

The citation period of a single publication is divided into two equal parts. Suppose

$t_h$ as the half of the citation period and $c_h$ as the number of citations received in the year $t_h$. $t_h$ is generally > 5, because delayed recognition papers are initially rarely cited over a period of 5 years but become highly cited during the next 15 years (see Glänzel & Garfield, 2004). $c_t$ is defined as the number of citations received in the $t^{th}$ year after publication, $t$ as the number of years since publication, and $c_0$ as the number of citations in the publication year. In the two equal parts of the citation period, let $t_1$ be the year in which the citations reach an early peak and $c_1$ the citations in the year $t_1 < t_h$. $t_2$ is defined as the year with the late citation peak and $c_2$ as the citations in the year $t_2 > t_h$. If $c_1$>10 and $t_1 \leq 5$, $c_1/t_1 > 2$, which lead to $\beta_1 >$ arctan(2) ~ 63°27'. Thus, $\beta_1 > 60°$ may identify SGs. $c_2$>10 and $t_2 \leq 15$ lead to $\beta_2 >$ arctan(2/3) ~ 33°41'. In case of $c_2$ >10 and $t_2 \leq 100$, $c_2/t_2 > 0.1$, $\beta_2$ = arctan(0.1) ~ 5°42', so that $\beta_2 > 5°$ may identify SBs.

According to van Raan (2004), SBs receive more than 20 citations after their awakening (within 4 years, including the peak year), which means that the 4 year citations (using symbol SCb) > 20. Then, SGs are suggested to receive more than 20 citations until an early peak (within 4 years, including the peak year), which means that the 4 year citations (using symbol SCa) > 20. The mean annual citations (AC) should be less than or equal to 2 for SBs during the sleeping period; the AC are suggested to be less than or equal to 10 for SGs after the peak.

We start from the premise that there are many papers in citation databases for which $t_1 \leq 5$ and $t_2 - t_1 \geq 10$. Thus, we suggest identifying SBs and SGs in the databases with the criteria listed in Table 1. In the table, we distinguish between possible, typical, and higher than typical SGs and SBs.

If an article shows typical characteristics of both SGs and SBs, we classify it as a typical ASB.

Table 1. Criteria for the identification of "smart girls" (SGs) and "sleeping beauties" (SBs)

| | $\beta_1$ | $\beta_2$ | Ca | Cb | AC | $t_2$-$t_1$ |
|---|---|---|---|---|---|---|
| **"Smart Girl" (SG)** | | | | | | |
| Possible SG | > 60° | | > 20 | | | |
| Typical SG | > 60° | | > 20 | | ≤ 10 | ≥ 10 |
| Higher than typical SG | > 88° | | > 20 | | ≤ 10 | ≥ 10 |
| **"Sleeping beauty" (SB)** | | | | | | |
| Possible SB | | > 5° | | > 20 | | |
| Typical SB | | > 5° | | > 20 | ≤ 2 | ≥ 10 |
| Higher than typical SB | | > 30° | | > 20 | ≤ 2 | ≥ 10 |

Because the beauty coefficient B (Ke et al., 2015) considers only one citation peak, it cannot be used to differentiate SGs from SBs. However, this is possible for the citation angle β. Visually, $c_1 \gg c_2$ is a tendency towards SGs and $c_1 \ll c_2$ a tendency towards SBs. $\beta_1 \gg \beta_2$ probably leads to SGs; whereas the probability of SBs increases with $\beta_1 \ll \beta_2$. The advantage of the citation angle $\beta$ lies in its ability to identify both SGs and SBs in a simple and straightforward way.

### 3.3 Dataset used

In order to exemplify the identification of SGs and SBs in a large dataset, we used all articles published in 1980 in the natural sciences (n=166,870). The data was derived from an in-house database at the Max Planck Society which is based on the Web of Science (WoS). We used the assignments of WoS journal sets to the broad subject area

"natural sciences" (undertaken by the OECD)[1] to select the natural sciences papers. For each article in the dataset, the annual number of citations is available from 1980 until 2015 (36 annual citation counts per paper). We restricted the dataset to papers with the document type "article", the publication year 1980, and the subject area "natural sciences" in order to compare the annual citation impact of papers which are (more or less) similar.

The analyses of this study are based on annual citations including self-citations. It is not possible to identify self-citations in our in-house database. This might be a limitation of the study, since self-citations can play a role in the awakening process of SBs (van Raan, 2015).

## 4. Results

In our dataset (n=166,870 papers), we identified 24,131 papers (14.5%) with a "smart" angle > 60$^o$ and 42,212 papers (25.3%) with a "sleeping" angle > 5$^o$. There are 6,800 papers (4.1%) with a "smart" angle > 60$^o$ and with SCa > 20; 3,456 papers (2.1%) have a "sleeping" angle > 5$^o$ with SCb > 20. Thus, about 4% of the papers are possible SGs and 2% possible SBs (according to the criteria in Table 1). Furthermore, we identified 5,718 papers (3.43%) in the dataset with a "smart" angle > 60$^o$, SCa > 20, and AC ≤ 10 (in $t_2$-$t_1$ ≥ 10). 126 papers (0.0755% ~ 0.1%) have a "sleeping" angle > 5$^o$ with SCb > 20 and AC ≤ 2 (in $t_2$-$t_1$ ≥ 10). Only 10 SGs and 7 SBs are higher than typical

1 http://ipscience-help.thomsonreuters.com/incitesLive/globalComparisonsGroup/globalComparisons/subjAreaSchemesGroup/oecd.html

cases. This means that about 3% of the papers are typical SGs, and about 0.1% typical SBs. As expected, SGs are a more frequent phenomenon than SBs.

Table 2. Main parameters of typical "all-elements-sleeping-beauties" (ASBs) as well as higher than typical "smart girls" (SGs) and "sleeping beauties" (SBs)

| Paper | WoS Accession number | Category/Categories | $\beta_1$ (º) | $\beta_2$ (º) | B |
|---|---|---|---|---|---|
| **Higher than typical SGs** | | | | | |
| Steinbacher, R. (1980). PRESELECTION OF SEX - THE SOCIAL-CONSEQUENCES OF CHOICE. *Sciences-New York, 20*(4), 6-9. | WOS:A1980JL87600005 | Multidisciplinary Sciences | 88.8983 | 0 | 0 |
| Vanallen, J. A., Thomsen, M. F., Randall, B. A., Rairden, R. L., & Grosskreutz, C. L. (1980). SATURNS MAGNETOSPHERE, RINGS, AND INNER SATELLITES. *Science, 207*(4429), 415-421. | WOS:A1980JB22600022 | Multidisciplinary Sciences | 88.0251 | 6.5819 | 0 |
| Cory, S., & Adams, J. M. (1980). DELETIONS ARE ASSOCIATED WITH SOMATIC REARRANGEMENT OF IMMUNOGLOBULIN HEAVY-CHAIN GENES. *Cell, 19*(1), 37-51. | WOS:A1980JC60800004 | Biochemistry & Molecular Biology; Cell Biology | 88.3634 | 5.7106 | 0 |
| Rich, A. (1980). BITS OF LIFE. *Sciences-New York, 20*(8), 10-&. | WOS:A1980KK00700003 | Multidisciplinary Sciences | 88.8983 | 2.0454 | 0 |
| Bethe, H. A. (1980). THE LIVES OF THE STARS. *Sciences-New York, 20*(8), 6-9. | WOS:A1980KK00700002 | Multidisciplinary Sciences | 88.8983 | 0 | 0 |
| Lancaster, J. B., & Whitten, P. (1980). FAMILY MATTERS. *Sciences-New York, 20*(1), 10-15. | WOS:A1980HY48600005 | Multidisciplinary Sciences | 88.8983 | 0 | 0 |
| Jonas, S. (1980). RX FOR HEALTH-CARE-DELIVERY - WHATS AILING OUR SYSTEM. *Environment, 22*(2), 14-&. | WOS:A1980JK35900006 | Environmental Sciences; Environmental Studies | 88.8983 | 0 | 0 |
| Lewis, H. W. (1980). SAFETY OF FISSION REACTORS. *Scientific American, 242*(3), 53-65. | WOS:A1980JF37200002 | Multidisciplinary Sciences | 88.9191 | 2.8624 | 0 |
| Weiner, J. (1980). PRIME TIME SCIENCE. *Sciences-New York, 20*(7), 6-11. | WOS:A1980KD49300007 | Multidisciplinary Sciences | 88.8983 | 0 | 0 |
| Tepperman, B. L., & Evered, M. D. (1980). GASTRIN INJECTED INTO THE LATERAL HYPOTHALAMUS STIMULATES SECRETION OF GASTRIC-ACID IN RATS. *Science, 209*(4461), 1142-1143. | WOS:A1980KE67300028 | Multidisciplinary Sciences | 88.8983 | 2.4896 | 0 |
| **Higher than typical SBs** | | | | | |
| Carlson, N. W., Jackson, D. J., Schawlow, A. L., Gross, M., & Haroche, S. (1980). SUPER-RADIANCE TRIGGERING SPECTROSCOPY. *Optics Communications, 32*(2), 350-354. | WOS:A1980JK05300034 | Optics | 36.8699 | 30.5792 | 84.0227 |
| Pfisterer, M., & Nagorsen, G. (1980). ON THE STRUCTURE OF TERNARY ARSENIDES. *Zeitschrift Fur Naturforschung Section B-a Journal of Chemical Sciences, 35*(6), 703-704. | WOS:A1980KA96400011 | Chemistry, Inorganic & Nuclear; Organic | 33.6901 | 31.8274 | 214.643 |
| Smirnov, G. M., & Mahan, J. E. (1980). DISTRIBUTED SERIES RESISTANCE IN PHOTO-VOLTAIC DEVICES - INTENSITY AND LOADING EFFECTS. *Solid-State Electronics, 23*(10), 1055-1058. | WOS:A1980KJ33200010 | Engineering, Electrical & Electronic; Physics, Applied; Physics, Condensed Matter | 30.9638 | 36.5289 | 233.641 |
| Vilenkin, A. (1980). EQUILIBRIUM PARITY-VIOLATING CURRENT IN A MAGNETIC-FIELD. *Physical Review D, 22*(12), 3080-3084. | WOS:A1980KU96000018 | Astronomy & Astrophysics; Physics, Particles & Fields | 0 | 34.7778 | 335.318 |
| Varotsos, P., & Alexopoulos, K. (1980). MIGRATION ENTROPY FOR THE BOUND FLUORINE MOTION IN ALKALINE-EARTH FLUORIDES. *Journal of Physics and Chemistry of Solids, 41*(5), 443-446. | WOS:A1980JR80000005 | Chemistry, Multidisciplinary; Physics, Condensed Matter | 33.6901 | 38.6598 | 300.586 |
| Kodama, H. (1980). CONSERVED ENERGY FLUX FOR THE SPHERICALLY SYMMETRIC SYSTEM AND THE BACK-REACTION PROBLEM IN THE BLACK-HOLE EVAPORATION. *Progress of Theoretical Physics, 63*(4), 1217-1228. | WOS:A1980JR85100012 | Physics, Multidisciplinary | 45 | 34.1145 | 237.728 |
| Heymann, D. L., Weisfeld, J. S., Webb, P. A., Johnson, K. M., Cairns, T., & Berquist, H. (1980). EBOLA HEMORRHAGIC-FEVER - TANDALA, ZAIRE, 1977-1978. *Journal of Infectious Diseases, 142*(3), 372-376. | WOS:A1980KN37600010 | Immunology; Infectious Diseases; Microbiology | 53.1301 | 40.3645 | 100.662 |
| **Typical ASBs** | | | | | |
| Fujita, T., Suzuki, M., Komatsubara, T., Kunii, S., Kasuya, T., & Ohtsuka, T. (1980). ANOMALOUS SPECIFIC-HEAT OF CEB6. Solid State Communications, 35(7), 569-57 | WOS:A1980KD49700015 | Physics, Condensed Matter | 66.8014 | 26.5651 | 7.8333 |
| Smith, E. J., Davis, L., Jones, D. E., Coleman, P. J., Colburn, D. S., Dyal, P., & Sonett, C. P. (1980). SATURNS MAGNETOSPHERE AND ITS INTERACTION WITH THE SOLAR-WIND. *Journal of Geophysical Research-Space Physics, 85*(NA11), 5655-5674. | WOS:A1980KS08700002 | Astronomy & Astrophysics | 77.4712 | 20.3231 | 0.3247 |
| Collins, M. D., Goodfellow, M., & Minnikin, D. E. (1980). FATTY-ACID, ISOPRENOID QUINONE AND POLAR LIPID-COMPOSITION IN THE CLASSIFICATION OF CURTOBACTERIUM AND RELATED TAXA. *Journal of General Microbiology*, 118(MAY), 29-37. | WOS:A1980JX85400004 | Microbiology | 74.0546 | 18.9246 | 0.2121 |
| Frank, L. A., Burek, B. G., Ackerson, K. L., Wolfe, J. H., & Mihalov, J. D. (1980). PLASMAS IN SATURNS MAGNETOSPHERE. *Journal of Geophysical Research-Space Physics, 85*(NA11), 5695-5708. | WOS:A1980KS08700005 | Astronomy & Astrophysics | 78.6901 | 19.0935 | 2.7524 |
| Wagner, F. T., & Somorjai, G. A. (1980). PHOTOCATALYTIC HYDROGEN-PRODUCTION FROM WATER ON PT-FREE SRTIO3 IN ALKALI HYDROXIDE SOLUTIONS. *Nature, 285*(5766), 559-560. | WOS:A1980JW42800027 | Multidisciplinary Sciences | 74.7449 | 13.627 | -0.333 |

Accordingly, there are 5 typical ASBs in the dataset which meet the criteria "smart" angle > 60°, "sleeping" angle > 5°, SCa > 20, SCb > 20, and AC ≤ 2 (in $t_2$-$t_1$ ≥ 10). The main parameters ($\beta_1$, $\beta_2$, and B) and bibliographic information of the typical ASBs as well as the higher than typical SGs and SBs are shown in Table 2. It is interesting to see that typical ASBs and higher than typical SBs in Table 2 are mostly published in field-specific journals. In contrast, the higher than typical SGs appeared in journals which are categorized as multidisciplinary by Thomson Reuters (e.g. *Science*). It seems that these journals focus on the publication of manuscripts addressing hot topics without long-term impact.

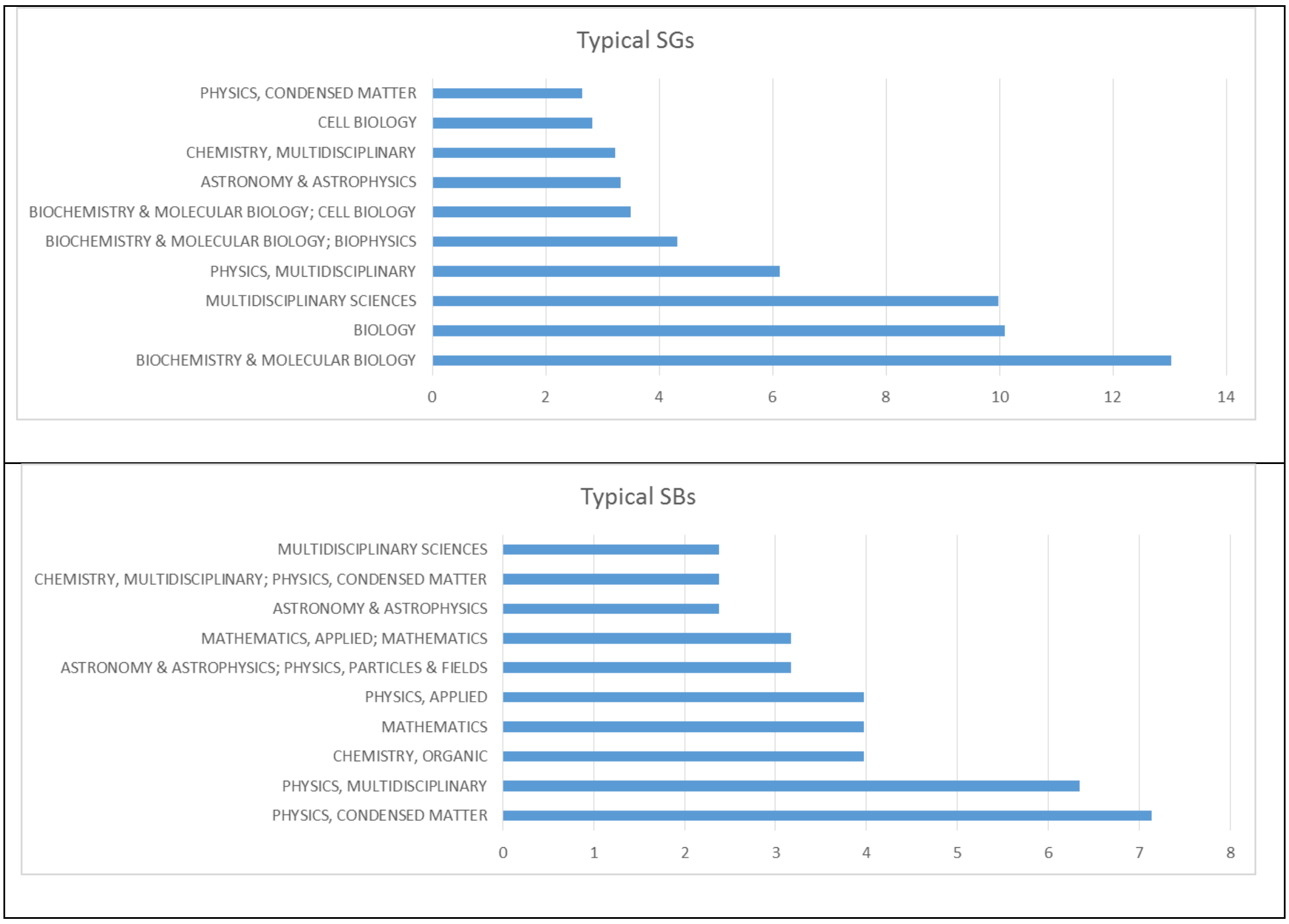

Figure 2. Percentage of typical "smart girls" (SGs) and sleeping beauties" (SBs) in different subject categories. The subject categories with the most SGs and SBs are shown. If papers belong to more than one category, the papers in the mixed categories are counted.

If we analyze the subject categories of the typical SGs (n=5718) and SBs (n=126) in our dataset, we receive the percentages in Figure 2. The results in the figure indicate that typical SGs are (with 10%) significantly more often published in journals of the multi-disciplinary category than typical SBs (with 2.4%).

In order to show typical citation curves of SGs, SBs, and ASBs, we show in the following two higher than typical cases of SGs and SBs as well as two typical cases of ASBs from Table 2. The selected six papers are shaded grey in the table. The higher than typical SG papers on the “preselection of sex – the social consequences of choice” (WOS:A1980JL87600005) and on “Saturn’s magnetosphere, rings, and inner satellites” (WOS:A1980JB22600022) exhibit a high citation impact at the beginning of their citation histories followed by early decreases. The further histories are characterized by zero or low citation counts with some peaks which do not, however, reach the level of the early phase. The papers have the following parameters: $\beta_1 = 88.9^o$, $\beta_2 = 0$, and B=0 (WOS: A1980JL87600005) and $\beta_1 = 88.02^o$, $\beta_2 = 6.58^o$, and B=0 (WOS: A1980JB22600022).

The selected higher than typical SBs from Table 2 are papers on the “super-radiance triggering spectroscopy” (WOS:A1980JK05300034) and “on the structure of ternary arsenides” (WOS:A1980KA96400011) with the following parameters .$\beta_1 = 36.87^o$, $\beta_2 = 30.58^o$, and B=84.02 (WOS:A1980JK05300034) and $\beta_1 = 33.39^o$, $\beta_2 = 31.83^o$, and B=214.64 (WOS:A1980KA96400011). The paper on the super-radiance triggering spectroscopy shows very low citation counts for nearly 20 years followed by a large increase around the year 2000 (see Figure 3). The paper on the structure of ternary arsenides shows a similar annual citation curve where the awakening of the paper happens at a later time point: nearly thirty years after its publication, the paper receives comparably high numbers of citations.

Figure 3 also shows two typical mixed cases which are both SGs and SBs. We selected the following two papers from Table 2 for visualizing their citation curves: the first paper is on "anomalous specific heat of CEB6" (WOS:A1980KD49700015) and the second on "Saturn's magnetosphere and its interaction with the solar wind" (WOS:A1980KS08700002). The corresponding parameters are $\beta_1$ = 66.8$^o$, $\beta_2$ =26.56$^o$, and B=7.83 (WOS:A1980KD49700015) and $\beta_1$ = 77.47$^o$, $\beta_2$ =20.32$^o$, and B=0.32 (WOS:A1980KS08700002). Both papers show an early peak in their citation curve with a subsequent decrease. A longer phase of years with low citation counts is finished by an impact increase 15 to 20 years after their publication.

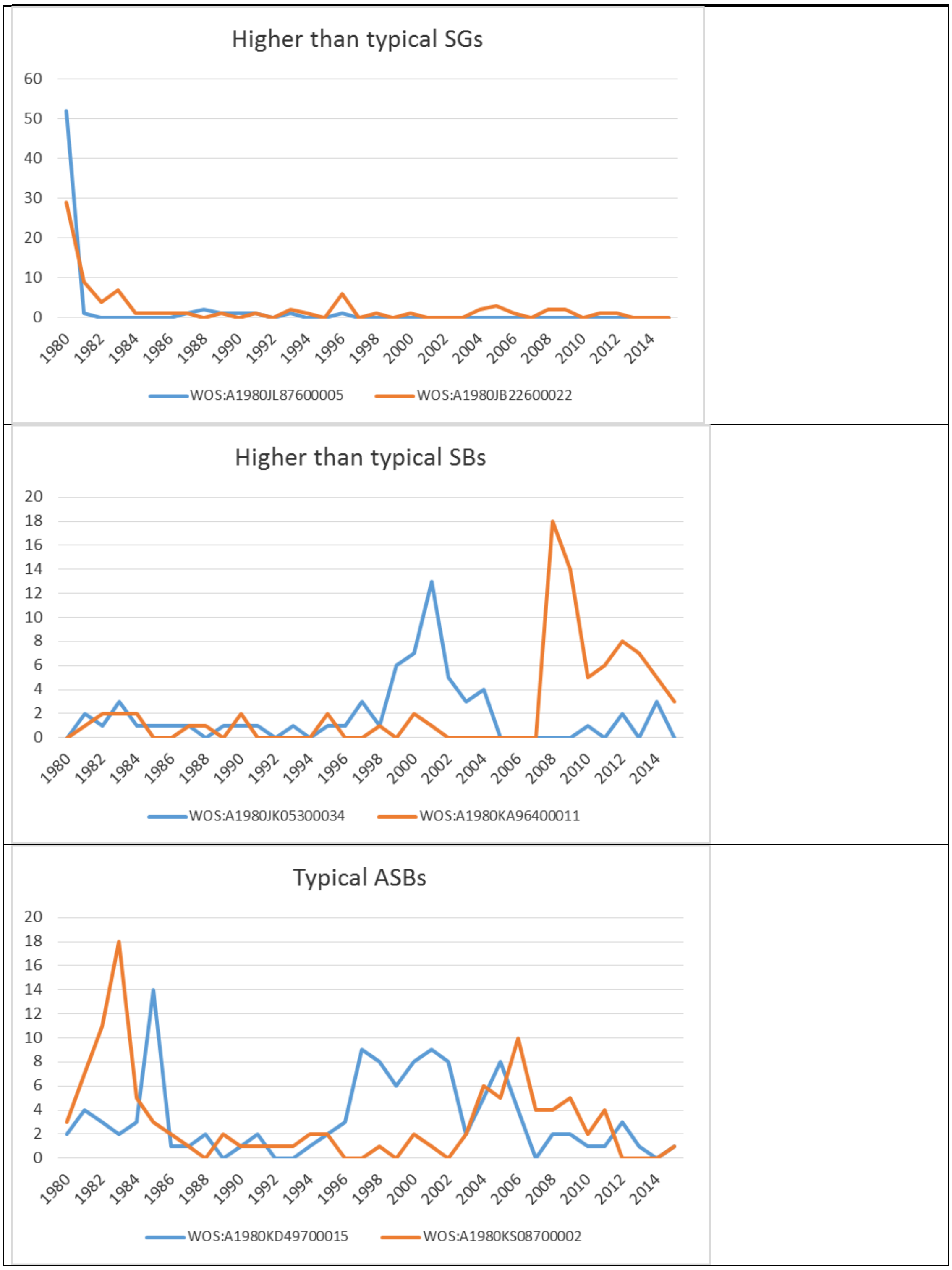

Figure 3. Annual citation curves of two higher than typical "smart girls" (SGs) and sleeping beauties (SBs) as well as two typical all-elements-sleeping beauties (ASBs)

## 5. Analysis and discussion

In the dynamic view, we can assume a wave-like annual citation curve in the period from 0 to *t* for most of the publications (see our examples in section 4). Figure 4 shows the dynamic view of a citation curve, where the first citation peak is located at $c_1$ and the late peak at $c_2$.

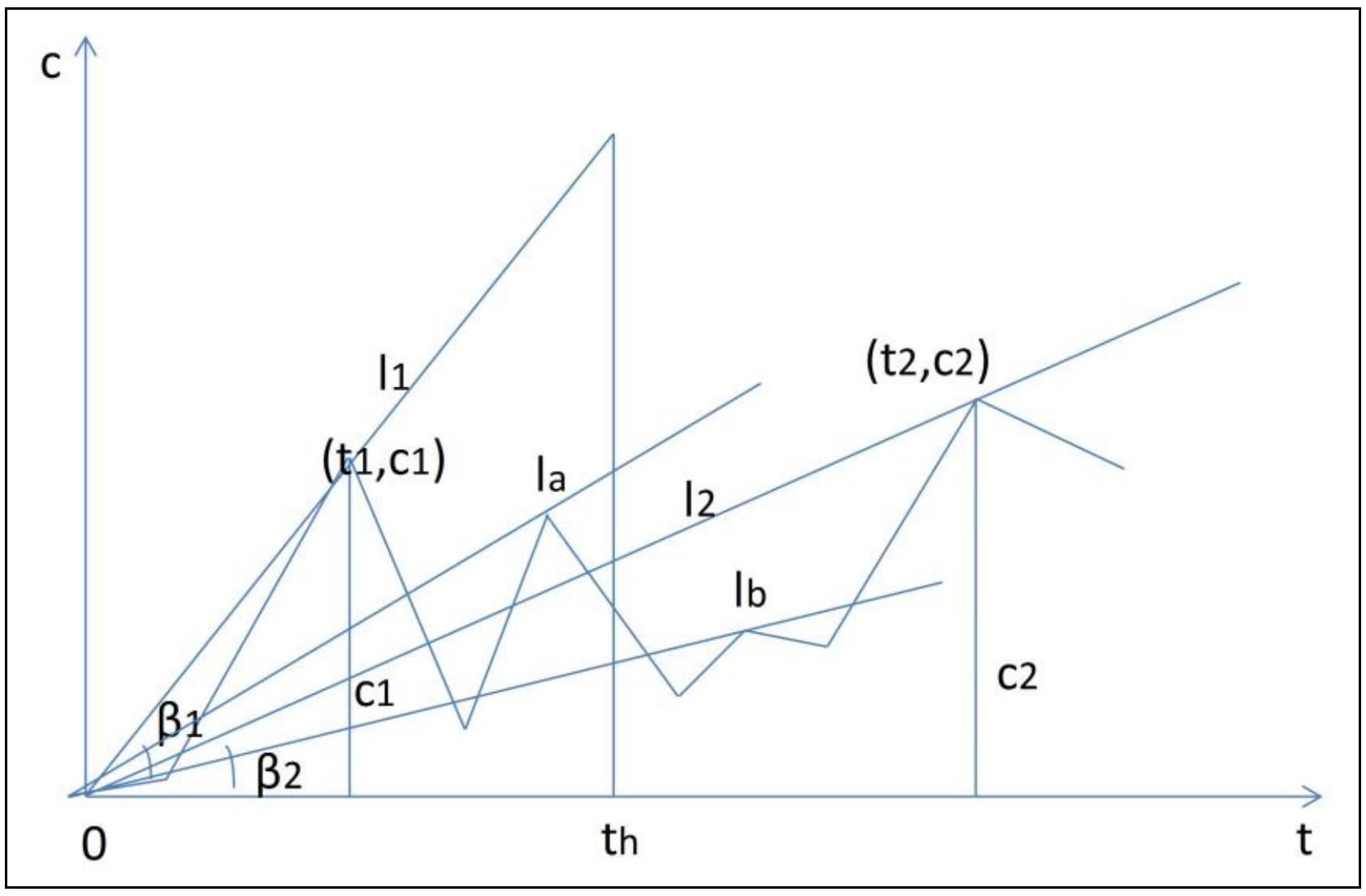

Figure 4. A dynamic view on a citation curve

Over the whole citation curve, the line which links the zero point with the citation peak extends dynamically via $l_1 \rightarrow l_a \rightarrow l_b \rightarrow l_2$, while the citation angle β changes from $\beta_1$ to $\beta_2$. Generally, given the same number of annual citations over time, the citation angle $\beta$ will decrease. When the citation curve reaches a late pronounced peak, $\beta$ becomes $\beta_2$. The citation angle $\beta$ is a simple indicator for measuring the change process in citation distributions, which could be used as a simple dynamic coefficient. Although the beauty coefficient B could also be used here, it does not show regular changes, because there is no simple regulation in its definition (c.f. Eq. 1).

Analytically, the citation angle $\beta$ is determined by two variables, annual citations $c$ and time $t$. Thus, $\beta$ is a function of $c$ and $t$

$$\beta(c,t) = \arctan(\frac{c}{t}) \tag{8}$$

Its change with $c$ and $t$ is

$$\frac{\partial\beta}{\partial c} = (\frac{1}{1+(c/t)^2})\frac{1}{t} = \frac{t}{t^2+c^2} \tag{9}$$

$$\frac{\partial\beta}{\partial t} = (\frac{1}{1+(\mathrm{c/t})^2})(-\frac{c}{t^2}) = -\frac{c}{t^2+c^2} \tag{10}$$

$$d\beta = \frac{\partial\beta}{\partial c}dc + \frac{\partial\beta}{dt}dt = \frac{tdc-cdt}{t^2+c^2} \tag{11}$$

We can see that changes of citation angle $\beta$ link not only annual citations $c$ with time $t$ themselves, but also their changes.

However, some limitations exist if the citation angle $\beta$ is used as a coefficient. As the angles are not mathematically sensitive, they may be an insensitive indicator. The change of the numerical value of a trigonometric function is not uniform. For $tan\beta$, its numerical value changes slightly for small angles. However, the numerical value of $tan\beta$ changes significantly and quickly for large angles approaching $90^o$.

Perhaps the citation angle $\beta$ could be combined with the beauty coefficient B. The combined measure may be used as a dynamic measure of annual citation curves. While the citation angle $\beta$ measures dynamic angles of change, the beauty coefficient B provides absolute values of the change. By using the citation angle $\beta$ and the beauty coefficient B as unified measures, the citation phenomena SG and SB can be simply united and uniformly studied.

SG and SB are two typical citation impact phenomena, which provide ways to explore the citation patterns of single publications. As SG and SB have typical meanings in various citation patterns (Avramescu, 1979; Li & Ye, 2014), they could act as starting points for the understanding of complex citation patterns. A first attempt is the consideration of ASBs in this study. Our results show that less than 5% of the articles from 1980 are SGs or SBs. Thus, more than 90% of the papers are neither SGs nor SBs. Further studies could try to identify additional typical citation patterns in the annual citation counts of papers.

## 5. Conclusions

Since bibliometrics is most frequently used in an evaluative context, the early citations of papers are normally the focus of interest. In evaluative bibliometrics, it is conventional practice to measure the citation impact of publications over a 3- to 5-year citation window. However, in recent years the so called SBs have received considerable attention from bibliometricians and beyond. People are fascinated by the few papers which receive only a few or no citations over a long period, before they start a late career. Bibliometricians speculate in case studies, why single publications are without any interest to the community over a long period and generate a lot of citation impact a long time after publication. For example, Marx (2014) analyzed the Shockley-Queisser paper as a notable example of an SB.

After the SG phenomenon was suggested in this paper as a synonym of instant credit or “flashes in the pan”, SGs and SBs were empirically unified and studied.

Against the backdrop of other approaches for identifying SBs, we introduced the citation angle β for measuring different SG and SB phenomena (typical and higher than typical cases). We showed that the citation angle β provides a simple indicator to differentiate SGs and SBs in annual citation distributions. While an early citation angle > 60° with SCa > 20 and AC ≤ 10 in $t_2-t_1 \geq 10$ characterizes typical SGs, a late citation angle > 5° with SCb > 20 and AC ≤ 2 in $t_2-t_1 \geq 10$ identifies typical SBs.

Despite its inherent limitations (see above), the citation angle $\beta$ is a promising measure to analyze and describe the changes in annual citation curves. The empirical analyses in this study demonstrate its practical usefulness. Since more or less pronounced early peaks (SGs) and late peaks (SBs) are characteristics of nearly all annual citation distributions, the citation angle β could be used to explore various citation patterns. The investigation of these patterns using the citation angle concept could be interesting topics for future studies.

**Acknowledgements**

The bibliometric data used in this paper are from an in-house database developed and maintained by the Max Planck Digital Library (MPDL, Munich) and derived from the Science Citation Index Expanded (SCI-E), Social Sciences Citation Index (SSCI), and Arts and Humanities Citation Index (AHCI) prepared by Thomson Reuters (Philadelphia, Pennsylvania, USA). We acknowledge the National Natural Science Foundation of China Grant No. 71673131. We thank Simon S. Li for program coding, and Qing Ke and Ludo Waltman, as well as two anonymous reviewers for helpful comments on earlier versions of this paper.